\documentclass[prd,preprint,showpacs,groupedaddress]{revtex4-1}
\usepackage{amsmath}
\usepackage{amsfonts}
\usepackage{amssymb}
\usepackage{geometry}
\usepackage{graphicx}
\usepackage{natbib}
\usepackage[english]{babel}
\usepackage{graphicx}
\usepackage{epstopdf}
\usepackage{subfigure}
\usepackage{caption}
\usepackage{multirow}
\usepackage{indentfirst}
\usepackage{diagbox}
\usepackage{mathrsfs}
\usepackage{booktabs}
\usepackage[colorlinks,linkcolor=blue,anchorcolor=,citecolor=blue]{hyperref}
\usepackage{amsthm}
\usepackage{float}

\begin{document}
    \title{Thermodynamics of AdS-Schwarzschild-like black hole in loop quantum gravity}
    
    \author{Rui-Bo Wang}
    \affiliation{Lanzhou Center for Theoretical Physics, Key Laboratory of Theoretical Physics of Gansu Province, Lanzhou University, Lanzhou, Gansu 730000, China}
    
    \author{Shi-Jie Ma}
    \affiliation{Lanzhou Center for Theoretical Physics, Key Laboratory of Theoretical Physics of Gansu Province, Lanzhou University, Lanzhou, Gansu 730000, China}
    
    \author{Lei You}
    \affiliation{Lanzhou Center for Theoretical Physics, Key Laboratory of Theoretical Physics of Gansu Province, Lanzhou University, Lanzhou, Gansu 730000, China}
    
    \author{Yu-Cheng Tang}
    \affiliation{Lanzhou Center for Theoretical Physics, Key Laboratory of Theoretical Physics of Gansu Province, Lanzhou University, Lanzhou, Gansu 730000, China}
    
    \author{Yu-Hang Feng}
    \affiliation{Lanzhou Center for Theoretical Physics, Key Laboratory of Theoretical Physics of Gansu Province, Lanzhou University, Lanzhou, Gansu 730000, China}
    
	\author{Xian-Ru Hu}
	\affiliation{Lanzhou Center for Theoretical Physics, Key Laboratory of Theoretical Physics of Gansu Province, Lanzhou University, Lanzhou, Gansu 730000, China}

	\author{Jian-Bo Deng}
	\email[Email: ]{dengjb@lzu.edu.cn}
	\affiliation{Lanzhou Center for Theoretical Physics, Key Laboratory of Theoretical Physics of Gansu Province, Lanzhou University, Lanzhou, Gansu 730000, China}
	\date{\today}
	
    \begin{abstract}
    We obtained the metric of the Schwarzschild-like black hole with loop quantum gravity (LQG) corrections in anti-de Sitter (AdS) space-time, under the assumption that the cosmological constant is decoupled in LQG. We investigated its thermodynamics, including the equation of state, criticality, heat capacity, and Gibbs free energy. The $P-v$ graph was plotted, and the critical behavior was calculated. It was found that, due to the LQG effect, the quantum-corrected Schwarzschild-AdS black hole exhibits a critical point and a critical ratio of $7/18$, which differs from the Reissner-Nordstr$\ddot{\mathrm{o}}$m-AdS black hole's ratio of $3/8$ (the same as that of the Van der Waals system) slightly. However, there are still some similarities compared to the Van der Waals system, such as the same critical exponents and a similar $P-v$ graph. Moreover, it is concluded that the energy-momentum tensor related to the black hole's mass could violate the conventional first law of thermodynamics. This modified first law may violate the conservation of Gibbs free energy during the small black hole-large black hole phase transitions, potentially indicating the occurrence of the zeroth-order phase transition. The Joule-Thomson expansion was also studied. Interestingly, compared to the Schwarzschild-AdS black hole, the LQG effect leads to inversion points. The inversion curve divides the $\left(P,T\right)$ coordinate system into two regions: a heating region and a cooling region, as shown in detail by the inversion curves and isenthalpic curves. The results indicated that there is a minimum inversion mass, below which any black hole will not possess an inversion point.
    \end{abstract}
   
    \maketitle
    \section{Introduction}\label{sec1}
    Since in 1983, Hawking and Page made their groundbreaking research on the thermodynamics of AdS black holes and discovered the phase transitions in Schwarzschild-AdS black holes~\cite{HawkingPage}, black hole thermodynamics has attracted widespread attention of researchers. In 1999, Chamblin discovered that there is the first-order phase transition in RN-AdS black holes, which is similar to the gas-liquid phase transition~\cite{RN1,RN2}. Kastor suggested that the mass of an AdS black hole should be regarded as its enthalpy, with the cosmological constant related to pressure, resulting in the first law of AdS black holes in an extended phase space~\cite{MPfirstlaw}. The equation of state for rotating and higher-dimensional black holes was studied in~\cite{Dolan}. In research~\cite{PVcri}, the authors introduced the criticality of charged-AdS black holes in detail. An interesting result is that the critical ratios ($P_{c}v_{c}/T_{c}$) of many black holes exhibit only a slight deviation from (or are even exactly equal to) the result of the Van der Waals system $3/8$~\cite{PVcri,QEDthermo1,highQEDthermo,maxarea1,maxarea2,ratio1,ratio2,ratio3,ratio4,ratio5,ratio6}, which exactly matches that of the Van der Waals system. However, this deviation can also be significant~\cite{Hayward,HoravaLifshitz,Reentrant,Geometrical,modifyfirst2,Kerr-Sen,ParametricGauss-Bonnet}. A notable finding is that M. M. Stetsko discovered in his work that the critical ratio for a four-dimensional static black hole in minimal Horndeski gravity with Maxwell and Yang-Mills fields is $75/512$, which is clearly less than $3/8$~\cite{Horndeski}. These anomalies may provide important implications for black hole thermodynamics and even for the studies of real gases.
    
    The theme of this paper is the thermodynamics of a type of quantum-corrected black holes. In comparison to classical Einstein gravity, quantum effects are widely considered by physicists. The theory that incorporates quantum electrodynamics (QED) and its implications for black holes have been researched in~\cite{QED1,QED2}. The thermodynamics of these black holes was studied in~\cite{QEDthermo1}, revealing that the QED parameter significantly influences phase transitions and criticality~\cite{QEDthermo2,QEDthermo3}. Moreover, the impact of high-order QED corrections on black hole phase transitions is also discussed~\cite{highQEDthermo}. On the other hand, in~\cite{oneloopSch}, the authors studied low-energy one-loop quantum corrections to the Schwarzschild geometry. In their expositions, general relativity is fundamentally a quantum field theory described by effective field theory, which suggests that low-energy degrees of freedom organize themselves as quantum fields governed by a local Lagrangian~\cite{EFT}. Loop quantum gravity (LQG) has received considerable attention. In LQG, space-time is discretized into a series of tiny units called loops, and the properties of black holes in this framework have been widely studied~\cite{LQG1,LQG2,LQG3,LQG4,LQG5,LQG6,LQG7,LQG8}. Recently, a quantum-corrected Schwarzschild space-time in LQG has been proposed~\cite{LQGmetric}. Interestingly, based on optical properties, the authors suggested that gravitational collapse in this model could be expressed as a process of white hole's generation~\cite{LQGmetric,LQGimage,LQGshadow}. In thermodynamics, black hole entropy in LQG was discussed in~\cite{LQGentropy}.
    
    In this paper, we intend to study the thermodynamics of Schwarzschild-like black holes in AdS space, including the thermodynamic first law, the equation of state and its criticality. The Joule-Thomson expansion will be discussed as well. Our work aims to enrich the thermodynamics of AdS black holes in LQG and provide a reference for theoretical studies or potential observations in the future. This article is organized as follows: In Sect.~\ref{sec2}, we obtain the metric of the AdS-Schwarzschild-like black hole in LQG, discuss the existence of black hole solutions. In Sect.~\ref{sec3}, we investigate the thermodynamics in detail, including the modified first law, the equation of state, behavior of the critical point, the heat capacity and the Gibbs free energy. In Sect.~\ref{sec4}, the Joule-Thomson process is researched. We calculate the inversion and show the inversion curves and the black hole's isenthalpic curves at length. Finally, we provide our conclusion and outlook in Sect.~\ref{sec5}. In this article, we use natural units $G=\hbar=k_{B}=c=1$.
    
 	\section{AdS-Schwarzschild-like black hole in LQG}\label{sec2}
    The metric of Schwarzschild-like black hole in LQG is~\cite{LQGmetric}
    \begin{equation}\label{eq2_1}
    	\mathrm{d}s^{2}=-f\left(r\right)\mathrm{d}t^{2}+\frac{1}{f\left(r\right)}\mathrm{d}r^{2}+r^{2}\mathrm{d}\theta^{2}+r^{2}\sin^{2}\theta \mathrm{d}\phi^{2},
    \end{equation}
    with
    \begin{equation}\label{eq2_2}
    	f\left(r\right)=1-\frac{2M}{r}+\frac{\alpha M^{2}}{r^{4}},
    \end{equation}
    where $\alpha=16\sqrt{3}\pi \gamma^{3}$, with $\gamma$ being the Barbero-Immirzi parameter in LQG~\cite{gamma}. $\alpha$ has a dimension of $\left[\mathrm{L}^{2}\right]$, and $\gamma$ is a dimensionless parameter. It should be noted that $\gamma$ is argued to be around $0.2$~\cite{gammavalue1,gammavalue2}, and interestingly, the calculation in~\cite{gammavalue1} suggests that $\gamma$ is approximately $0.2375$. This value has been widely used in~\cite{LQGmetric,LQGimage,LQGshadow,gammavalue3}. Therefore, in this article, we assume $\alpha$ to be positive and choose $\gamma$ as $0.2375$ or around $0.2$ to ensure the accuracy of the figures and table presented.
    
\begin{figure}[htbp]
    	\centering
    	\includegraphics[width=0.6\textwidth]{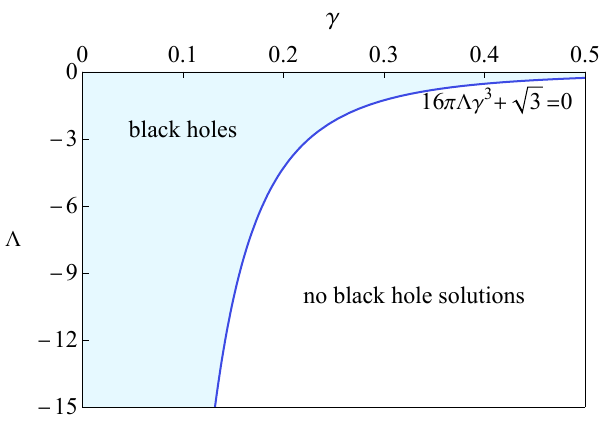}
    	\caption{The parameter space $\left(\gamma,\Lambda\right)$ distinguishes between black hole solutions and non-black hole solutions. The blue region represents the existence of black hole, while the white region indicates the absence of black hole solutions.}\label{pc}
    \end{figure}

    Assuming that the cosmological constant is decoupled in LQG theory. According to the metric generation approach discussed in~\cite{metric}, AdS-Schwarzschild-like black hole in LQG is described by
    \begin{equation}\label{eq2_3}
    	f\left(r\right)=1-\frac{2M}{r}+\frac{\alpha M^{2}}{r^{4}}-\frac{\Lambda r^{2}}{3}.
    \end{equation}
    The black hole solution requires $f\left(r_{+}\right)=0$, which leads to
    \begin{equation}\label{eq2_4}
    	M=\frac{3r_{+}^{3}\pm r_{+}^{2}\sqrt{-9\alpha+9r_{+}^{2}+3\alpha\Lambda r_{+}^{2}}}{3\alpha},
    \end{equation}
    where $r_{+}$ is the radius of event horizon. The positive branch is discarded because it can not converge to the Schwarzschild-AdS case when $\alpha \to 0$, but instead diverges to infinity. The black hole solution requires that both $M$ and $r_{+}$ are positive real numbers, which gives this following constraint,
    \begin{equation}\label{eq2_5}
    	\beta:=\frac{3+\alpha \Lambda}{3}>0,
    \end{equation}
    or a constraint on $\gamma$
    \begin{equation}\label{eq2_6}
    	16\pi\gamma^{3}\Lambda+\sqrt{3}>0.
    \end{equation}
    This condition is shown in Fig.~\ref{pc}.
    
    And there is a minimum radius of event horizon
    \begin{equation}\label{eq2_7}
    	r_{+}\geq\sqrt{\frac{\alpha}{\beta}}=r_{+}^{\mathrm{min}}.
    \end{equation}
    The negative branch of Eq.~\ref{eq2_4} with the condition Eq.~\ref{eq2_7} will give a minimum mass for the black hole:
    \begin{equation}\label{eq2_8}
    	M_{\mathrm{min}}=\sqrt{\alpha}\left(\frac{3-4\beta+\sqrt{9-8\beta}}{6\beta-6\beta^{2}}\right)^{\frac{3}{2}}\left(1-\sqrt{\frac{-3\beta+2\beta^{2}+\beta\sqrt{9-8\beta}}{3-4\beta+\sqrt{9-8\beta}}}\right),
    \end{equation}
    It could be verified that $\lim \limits_{\Lambda \to 0}M_{\mathrm{min}}=4\sqrt{\alpha}/3\sqrt{3}$, which agrees with the conclusion in~\cite{LQGmetric}. 
    
    \section{Thermodynamics}\label{sec3}
    \subsection{Thermodynamic functions and the first law}
    The black hole's temperature could be obtained from its surface gravity
    \begin{equation}\label{eq3_a1}
    	T=\frac{f'\left(r_{+}\right)}{4\pi}=\frac{ 2\alpha-3r_{+}^{2}-\alpha\Lambda r_{+}^2+\sqrt{-9 \alpha r_{+}^{2}+9r_{+}^{4}+3\alpha\Lambda r_{+}^{4}}}{2\pi\alpha r_{+}}.
    \end{equation}
    In the extended phase space, the cosmological constant is related with the pressure~\cite{MPfirstlaw}
    \begin{equation}\label{eq3_a2}
    	P=-\frac{\Lambda}{8\pi}.
    \end{equation}
    In black hole thermodynamics, the mass $M$ of the black hole is regarded as its enthalpy~\cite{MPfirstlaw}, and the first law can be expressed as follows:
    \begin{equation}\label{eq3_a3}
     	\mathrm{d}M=T \mathrm{d}S+V \mathrm{d}P.
    \end{equation}
    Using this, the entropy could be calculated as
    \begin{equation}\label{eq3_a4}
    	S=\int\left(\frac{\mathrm{d}M}{T}\right)_{P}=\frac{ \pi\sqrt{\beta^2r_{+}^{4}-\alpha\beta r_{+}^{2}}+\pi\alpha\ln\left(3\beta r_{+}+\sqrt{9\beta^2r_{+}^2-9\alpha\beta}\right)}{\beta \sqrt{\beta}}.
    \end{equation}
    Of note is that Bekenstein-Hawking area relation does not hold in this context. In fact, this type of energy-momentum tensor, which includes the black hole's mass $M$, will violate the conventional first law. This problem could be resolved by modifying the first law~\cite{modifyfirst1,modifyfirst2}. The modified first law is expressed as
    \begin{equation}\label{eq3_a5}
    	\mathrm{d}\mathcal{M}=T\mathrm{d}S+V\mathrm{d}P,
    \end{equation}
    where $\mathrm{d}\mathcal{M}=W\left(r_{+},\Lambda\right)\mathrm{d}M$, and $W$ is called correction function
    \begin{equation}\label{eq3_a6}
    	W=1+\int_{r_{+}}^{+\infty}4\pi r^{2} \frac{\partial T_{0}^{0}}{\partial M}\mathrm{d}r,
    \end{equation}
    where the energy-momentum tensor $T_{0}^{0}$ satisfies the Einstein equation
    \begin{equation}\label{eq3_a7}
    	R_{\mu}^{\nu}-\frac{1}{2}\delta_{\mu}^{\nu}R+\delta_{\mu}^{\nu}\Lambda=8\pi T_{\mu}^{\nu}.
    \end{equation} 
    Eq.~\ref{eq3_a6} is general and we will provide a brief proof in Appendix.~\ref{app1}. Specifically, for the black hole under discussion, $W$ is
    \begin{equation}\label{eq3_a8}
    	W\left(r_{+},\Lambda \right)=\sqrt{1-\frac{3\alpha-\alpha\Lambda r_{+}^{2}}{3r_{+}^{2}}}.
    \end{equation}
    Now the entrophy could be obtained by the new first law
    \begin{equation}\label{eq3_a9}
    	S=\int \left(\frac{\mathrm{d}\mathcal{M}}{T}\right)_{P}=\int\frac{1}{T}\left(\frac{\partial \mathcal{M}}{\partial r_{+}}\right)_{P}\mathrm{d}r_{+}=\pi r_{+}^{2},
    \end{equation} 
    which precisely accords with Bekenstein-Hawking function $S=A/4$. The thermodynamic volume of the black hole is
    \begin{equation}\label{eq3_a10}
    	V=\left(\frac{\partial \mathcal{M}}{\partial P}\right)_{S}=W\left(r_{+},\Lambda\right)\left(\frac{\partial M}{\partial P}\right)_{r_{+}}=\frac{4\pi r_{+}^{3}}{3}.
    \end{equation}
    One could demonstrate that the differential form $\mathrm{d}\mathcal{M}=T\mathrm{d}S+V\mathrm{d}P$ is not exact generally. In other words, it is impossible to define a scalar function $\mathcal{M}$ such that it satisfies the modified first law.
    
    \subsection{Equation of state and critical point}
\begin{figure}[htbp]
    	\centering
    	\includegraphics[width=0.6\textwidth]{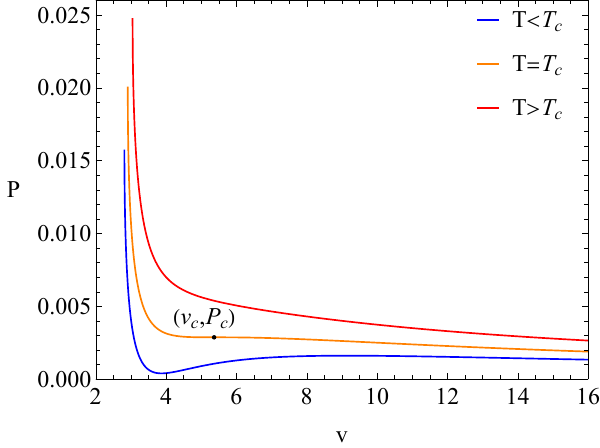}
    	\caption{The $P-v$ graph for the AdS-Schwarzschild-like black hole in LQG. The red, orange and blue lines correspond to $T=0.78T_{c}$, $T=T_{c}$ and $T=1.3T_{c}$. The black point on the orange line is the critical point. We set $\gamma=0.2375$.}\label{Pv}
    \end{figure}

    From Eq.~\ref{eq3_a1}, one could get the expression of $P$ as
    \begin{equation}\label{eq3_b1}
    	P=\frac{-4\alpha+4\pi\alpha  Tr_{+}+3r_{+}^{2}\pm\sqrt{-12\alpha r_{+}^{2}-24\pi\alpha Tr_{+}^{3}+9r_{+}^{4}}}{16\alpha \pi r_{+}^{2}}.
    \end{equation}
    One could find that the positive branch has no physical significance because it does not converge to the result of the Schwarzschild-AdS black hole if $\alpha \to 0$. Here, we introduce the specific volume $v=2r_{+}$~\cite{PVcri,v2r}, the equation of state is written as
    \begin{equation}\label{eq3_b2}
    	P=\frac{-16\alpha+8\pi\alpha T v+3v^{2}-\sqrt{-48\alpha v^{2}-48\pi\alpha T v^{3}+9v^{4}}}{16\alpha \pi v^{2}}.
    \end{equation}
    Now we investigate criticality. The critical point is obtained by
    \begin{equation}\label{eq3_b3}
    	\frac{\partial P}{\partial v}=0,\quad\frac{\partial^{2} P}{\partial v^{2}}=0,
    \end{equation}
    which leads to
    \begin{equation}\label{eq3_b4}
    	P_{c}=\frac{21}{2000\alpha \pi},\quad v_{c}=\frac{20\sqrt{5\alpha}}{9},\quad T_{c}=\frac{3 \sqrt{5}}{50 \pi \sqrt{\alpha}}.
    \end{equation}
    Compared to a normal Schwarzschild-AdS black hole, the LQG correction introduces a critical point. We take $\gamma=0.2375$ for an example and plot the $P-v$ graph in Fig.~\ref{Pv}.
    
    The figure reveals that the behavior of isothermal curve is similar to that of Van der Waals fluid. And it is necessary to point out that in the $P-v$ graph, $v$ cannot be arbitrarily small, as Eq.~\ref{eq2_7} and Eq.~\ref{eq3_b2} will yield a minimum value of $r_{+}$ for a fixed temperature $T$:
    \begin{equation}\label{eq3_b5}
    	r_{+}^{\mathrm{min}}=\frac{2}{3}\left(2\alpha\pi T+ \sqrt{3\alpha+4\alpha^{2}\pi^{2}T^{2}}\right).
    \end{equation}
    
    The critical ratio is
    \begin{equation}\label{eq3_b6}
    	\frac{P_{c}v_{c}}{T_{c}}=\frac{7}{18}.
    \end{equation}
    This result is different from that of the RN-AdS black hole, which is also the result of the Van der Waals system~\cite{PVcri}. An interesting phenomenon is that the critical ratios of many black holes show only a slight difference from $3/8$~\cite{PVcri,QEDthermo1,highQEDthermo,maxarea1,maxarea2,ratio1,ratio2,ratio3,ratio4,ratio5,ratio6}. However, this discrepancy also can be significant~\cite{Hayward,HoravaLifshitz,Reentrant,Geometrical,modifyfirst2,Kerr-Sen,ParametricGauss-Bonnet}. A typical example is found in the work of M. M. Stetsko, who discovered that the critical ratio for a four-dimensional static black hole in minimal Horndeski gravity with Maxwell and Yang-Mills fields is $75/512$, which is noticeably less than $3/8$~\cite{Horndeski}. These anomalies will provide valuable insights for the study of black hole thermodynamics and the equations of state for gases. Specifically, among the numerous data points, two results are particularly close to our findings. In~\cite{HoravaLifshitz}, the authors found that $z=4$ Ho\v{r}ava-Lifshitz black holes with hyperbolic horizons could exhibit two critical points, resulting in two critical ratios $0.463188$ and $0.386812$, the latter of which is close to $7/18$. Additionally, critical ratio of the Hayward-AdS black hole is $0.393031$, which is also very close to our result~\cite{Hayward}.
    
    Furthermore, defining
    \begin{equation}\label{eq3_b7}
    	p=\frac{P}{P_{c}},\quad \nu=\frac{v}{v_{c}},\quad \tau=\frac{T}{T_{c}},
    \end{equation}
    one could derive the equation of corresponding state
    \begin{equation}\label{eq3_b8}
    	p=\frac{125}{7}+\frac{9\tau}{7\nu}-\frac{27}{7\nu^{2}}-\frac{5\sqrt{5}}{7}\sqrt{125-\frac{18\tau}{\nu}-\frac{27}{\nu^{2}}}.
    \end{equation}
   	\subsection{Heat capacity and Gibbs free energy}
\begin{figure}[htbp]
   		\centering
   		\includegraphics[width=1\textwidth]{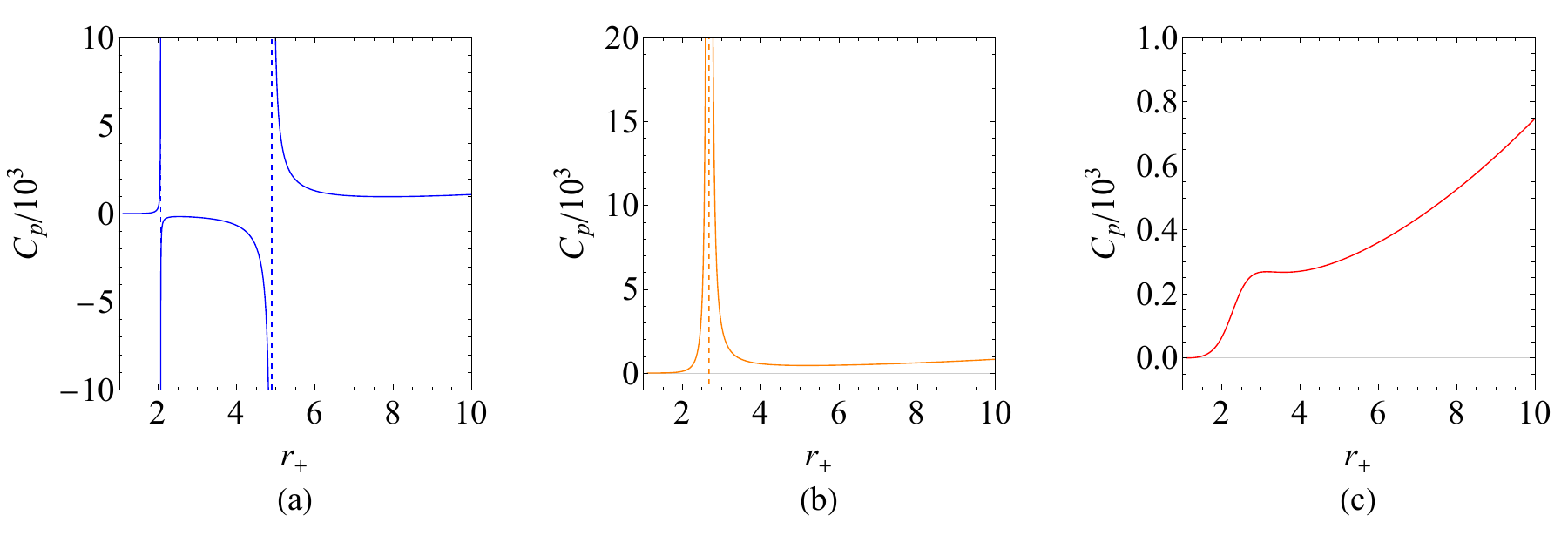}
   		\caption{The isobaric heat capacity of the black hole. We set $\gamma=0.2375$ and (a)$P=0.5P_{c}$; (b)$P=P_{c}$; (c)$P=1.5P_{c}$.}\label{CP1}
   	\end{figure}

   	The isobaric heat capacity is
   	\begin{equation}\label{eq3_c1}
   		C_{P}=T\left(\frac{\partial S}{\partial T}\right)_{P}=\frac{2 \pi r_{+}^{2}\sqrt{-3\alpha+3\beta r_{+}^{2}}\left( 3\beta r_{+}^{2}-2\alpha-\sqrt{9\beta r_{+}^{4}-9\alpha r_{+}^{2}}\right)}{-3\sqrt{3}\beta r_{+}^{3}+\left(2\alpha+3\beta r_{+}^{2}\right)\sqrt{3\beta r_{+}^{2}-3\alpha}}.
   	\end{equation}
   	The graph of the isobaric heat capacity is shown in Fig.~\ref{CP1}.
   	
   	As shown in the figure, the heat capacity diverge to infinity at the phase transition point, which is marked by the dashed lines. When $P<P_{c}$ (see Fig.~\ref{CP1}(a)), the small black holes-large black holes phase transition is similar to the first-order phase transition in traditional thermodynamics. At the critical point (see Fig.~\ref{CP1}(b)), the phase transition is similar to the second-order phase transition. We will reanalyze this phase transition later by calculating the Gibbs free energy.
   	
\begin{figure}[htbp]
   		\centering
   		\includegraphics[width=0.6\textwidth]{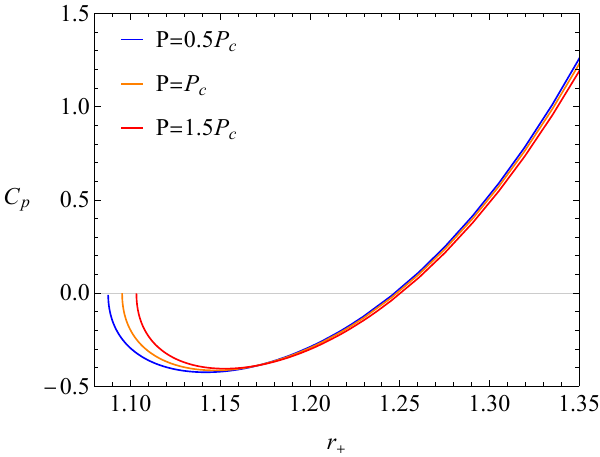}
   		\caption{The isobaric heat capacity will be negative if $r_{+}$ is too small. In the figure, $r_{+}\in[1.08,1.35]$ and $\gamma=0.2375$.}\label{CP2}
   	\end{figure}

   	It should be noticed that when $r_{+}$ is small,
   	\begin{equation}\label{eq3_c2}
   		r_{+}<\sqrt{\frac{3+4\alpha \Lambda-\sqrt{9-24\alpha \Lambda}}{2\Lambda\left(3+\alpha \Lambda\right)}},
   	\end{equation}
   	$C_{P}$ is negative. This means that a black hole with a small radius of the horizon is unstable. This negative region of $C_{P}$ is shown in Fig.~\ref{CP2}. 

\begin{figure}[htbp]
   		\centering
   		\includegraphics[width=0.6\textwidth]{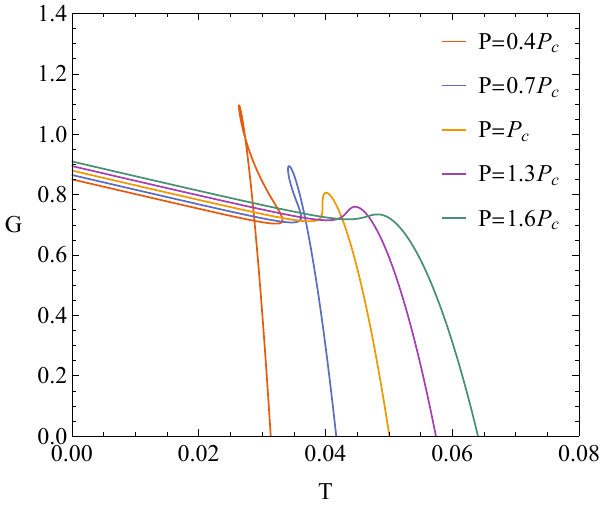}
   		\caption{The Gibbs free energy for different pressures. We set $\gamma=0.2375$. $P=0.4P_{c}$, $0.7P_{c}$, $P_{c}$, $1.3P_{c}$, $1.6P_{c}$ for five curves respectively.}\label{GT}
   	\end{figure}

\begin{figure}[htbp]
   		\centering
   		\includegraphics[width=0.6\textwidth]{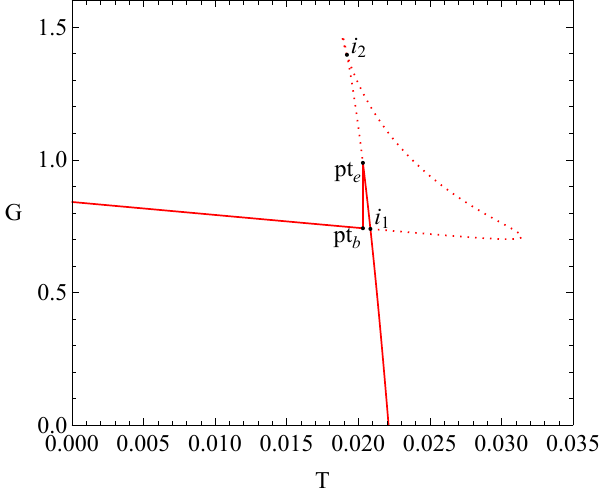}
   		\caption{The Gibbs free energy and the isothermal phase transition. We set $\gamma=0.2375$ and $P=0.2P_{c}$. Points pt$_{b}$ and pt$_{e}$ are the beginning and the end of the phase transition. Points $i_{1}$ and $i_{2}$ are two intersections of $G\left(T\right)$. The result after considering Maxwell's area law is marked by the solid line, while the other part is marked by the dotted line.}\label{dg}
   	\end{figure}

	Now we intend to investigate the Gibbs free energy. Although the differential form $-S\mathrm{d}T+V\mathrm{d}P$ is not exact because the first law has been modified, we still define the Gibbs free energy as follows:
	\begin{equation}\label{eq3c_3}
		G=M-TS=\frac{15r_{+}^{3}+3\alpha \Lambda r_{+}^{3}-6\alpha r_{+}-5r_{+}^{2}\sqrt{9r_{+}^{2}+3\alpha \Lambda r_{+}^{2}-9\alpha}}{6\alpha}.
	\end{equation} 
   	The Gibbs free energy is shown in Fig.~\ref{GT}.
   	The behavior of Gibbs free energy is similar to that of the Van der Waals system, $G$ is multivalued when phase transition occurs. But, there are also some differences. In fact, $G\left(T\right)$ is not necessarily a self-intersecting function even for $P<P_{c}$. Even if $G\left(T\right)$ has intersection points, these points do not indicate phase transition points. In the Van der Waals system, the first-order phase transition is an isothermal and isobaric process, and since $\mathrm{d}G=-S\mathrm{d}T+V\mathrm{d}P$ holds, the Gibbs free energy is conserved during the phase transition. But for this kind of black holes, the conventional first law does not work, and the differential form of $G$ becomes
   	\begin{equation}\label{eq3c_4}
   		\mathrm{d}G=-S\mathrm{d}T+W^{-1}V\mathrm{d}P+\left(W^{-1}-1\right)T\mathrm{d}S.
   	\end{equation} 
   	Even for an isothermal and isobaric process, $G$ is not conserved. The variation of Gibbs free energy during the phase transition is
   	\begin{equation}\label{eq3c_5}
   		\Delta G=\int_{S_{1}}^{S_{2}}\left(W^{-1}-1\right)T\mathrm{d}S,
   	\end{equation}
   	where $S_{1}$ and $S_{2}$ should be determined by Maxwell's area law in the $\left(S, T\right)$ phase space~\cite{QEDthermo1}
   	\begin{equation}\label{eq3c_6}
   		\begin{aligned}
   			\int_{S_1}^{S_2}T\mathrm{d}S=&T\left(S_{1}\right)\left(S_{2}-S_{1}\right),\\
   			T\left(S_{1}\right)&=T\left(S_{2}\right).
   		\end{aligned}
   	\end{equation}
   	Taking the $\gamma=0.2375, P=0.2P_{c}$ for an example, Eq.~\ref{eq3c_6} gives the beginning and the end of the phase transition in the $\left(T, G\right)$ coordinate:
   	\begin{equation}
   		\begin{aligned}
   			\mathrm{pt}_{b}&=\left(0.020328,0.74174\right),\\
   			\mathrm{pt}_{e}&=\left(0.020328,0.98797\right).
   		\end{aligned}
   	\end{equation}
   	But $G\left(T\right)$ has its self-intersections
   	\begin{equation}
   		\begin{aligned}
   			i_{1}&=\left(0.020852,0.73923\right),\\
   			i_{2}&=\left(0.019214,1.3949\right).
   		\end{aligned}
   	\end{equation}
   	These results are shown in Fig.~\ref{dg}.
   	
   	It is shown in the figure that, considering Maxwell's area law, the actual physical process could be described as follows: If the black hole undergoes an isobaric expansion, its temperature initially increase until it reaches the phase transition point pt$_{b}$, followed by an isothermal phase transition. During this process, the Gibbs free energy experiences a sudden change in $G\left(T\right)$ graph, rather than a continuous variation through the intersections $i_{1}$ and $i_{2}$. After the end of the phase transition pt$_{e}$, the temperature begins to rise again, and the Gibbs free energy decreases continuously. For our example with $\gamma=0.2375$ and $P=0.2P_{c}$, the change in Gibbs free energy during the phase transition is $\Delta G=0.24623\neq 0$.
   	
   	It should be noted that, according to Ehrenfest's classification of phase transitions, a discontinuity in the Gibbs free energy indicates a zeroth-order phase transition. Therefore, the phase transition from a small black hole to a large black hole that we are discussing should more accurately be referred to as a zeroth-order phase transition. The zeroth-order phase transition was first discovered in a dilatonic black hole~\cite{zeroth1}, and was also independently confirmed in the slowly rotating Einstein–Maxwell-dilaton black hole~\cite{zeroth2}. However, we would like to point out that there is a difference between the mechanisms of these two types of zeroth-order phase transitions. In studies~\cite{zeroth1,zeroth2}, the zeroth-order phase transitions observed by the researchers are induced by the dilaton-electromagnetic coupling interaction, and for these black holes, the first law of thermodynamics $\mathrm{d}M=T\mathrm{d}S+V\mathrm{d}P+\Phi\mathrm{d}Q$ still holds, where $\Phi$ is the electric potential and $Q$ is the electric charge. In contrast, for the Schwarzschild black holes with LQG corrections, the first law of thermodynamics is violated, and Eq.~\ref{eq3c_5} leads to the discontinuity of the Gibbs free energy, which is originally conserved during isothermal and isobaric phase transitions. Moreover, since Eq.~\ref{eq3c_5} is a general conclusion, for a four-dimensional static spherically symmetric black holes where the energy-momentum tensor explicitly depends on the black hole's mass, a zeroth-order phase transition is inevitable when $P<P_{c}$.
   	
   	\subsection{Critical exponents}
   	To calculate the critical exponents, we define
    \begin{equation}\label{eq3_d1}
    	t=\frac{T}{T_{c}}-1,\quad \omega=\frac{V}{V_{c}}-1,
    \end{equation}
    where $V_{c}$ is the thermodynamic volume corresponding to $v_{c}$. Near the critical point, one have $t\simeq 0$ and $\omega\simeq 0$, Eq.~\ref{eq3_b8} could be approximated as the series
    \begin{equation}\label{eq3_d2}
    	p=1+\frac{81}{28}t-\frac{135}{112}t\omega+\frac{81}{896}t^{2}-\frac{9}{140}\omega^{3}+\mathcal{O}\left(t\omega^{2},t^{2}\omega,\omega^{4}\right),
    \end{equation}
    which gives
    \begin{equation}\label{eq3_d3}
    	\left(\frac{\mathrm{d}p}{\mathrm{d}\omega}\right)_{T}=-\left(\frac{135}{112}t+\frac{27}{140}\omega^{2}\right).
    \end{equation}
    The definition of critical exponents $\epsilon$, $\lambda$, $\rho$ and $\delta$ could be found in~\cite{exponentdef}:
    \begin{equation}\label{eq3_d4}
     	C_{V}=T\left(\frac{\partial S}{\partial T}\right)_{V}\propto\left|t\right|^{-\epsilon},
    \end{equation}
    \begin{equation}\label{eq3_d5}
		\eta=V_{l}-V_{s}\propto\left|t\right|^{\lambda},  	
    \end{equation}
    \begin{equation}\label{eq3_d6}
		\kappa_{T}=-\frac{1}{V}\left(\frac{\partial V}{\partial P}\right)_{T}\propto\left|t\right|^{-\rho},
    \end{equation}
    \begin{equation}\label{eq3_d7}
    	\left|P-P_{c}\right|\propto\left|V-V_{c}\right|^{\delta},
    \end{equation}
    where $C_{V}$ is the isochoric heat capacity, $\eta$ is the change in thermodynamic volume during the phase transition (from a small black hole $V_{s}$ to a large $V_{l}$ black hole), $\kappa_{T}$ is the isothermal compressibility, and Eq.~\ref{eq3_d7} describes the behavior on the isothermal line $T=T_{c}$.
    
    It is evident that $C_{V}=0$ leads to $\epsilon=0$. $\eta$ could be obtained by Maxwell's area law in $\left(V,P\right)$ phase space~\cite{PVcri,modifyfirst2,maxarea1,maxarea2}:
    \begin{equation}\label{eq3_d8}
    	\oint V\mathrm{d}P=0,
    \end{equation}
    Using Eq.~\ref{eq3_d2} and Eq.~\ref{eq3_d3}, Maxwell's area law reads
    \begin{equation}\label{eq3_d9}
    	1+\frac{81}{28}t-\frac{135}{112}t\omega_{s}+\frac{81}{896}t^{2}-\frac{9}{140}\omega_{s}^{3}=1+\frac{81}{28}t-\frac{135}{112}t\omega_{l}+\frac{81}{896}t^{2}-\frac{9}{140}\omega_{l}^{3},
    \end{equation}
    \begin{equation}\label{eq3_d10}
    	\int_{\omega_{s}}^{\omega_{l}}-\left(\omega+1\right)\left(\frac{135}{112}t+\frac{27}{140}\omega^{2}\right)\mathrm{d}\omega=0,
    \end{equation}
    where $\omega_{s}$ and $\omega_{l}$ correspond to $V_{s}$ and $V_{l}$ respectively. The solution of this equation is $\omega_{l}=-\omega_{s}=5\sqrt{-3t}/2$, which gives $\lambda=1/2$. From Eq.~\ref{eq3_d3}, one have 
    \begin{equation}\label{eq3_d11}
    	\kappa_{T}=-\frac{1}{P_{c}(\omega+1)}\left(\frac{{\mathrm{d} \omega}}{\mathrm{d} p}\right)_{T}=\frac{112}{135P_{c}}\frac{1}{t}+\mathcal{O}\left(\omega\right),
    \end{equation}
    which indicates $\rho=1$. Finally, at the isothermal curve $T=T_{c}$, $t=0$ insures
    \begin{equation}\label{eq3_d12}
    	p=1-\frac{9}{140}\omega^{3}+\mathcal{O}\left(\omega^{4}\right).
    \end{equation}
    It gives the last critical exponent $\delta=3$. It is found that all the exponents are completely consistent with that of the Van der Waals fluid, the RN-AdS black hole and the Bardeen-Kiselev black hole~\cite{PVcri,modifyfirst2}. Naturally, it conforms to Griffiths, Rushbrooke and Widom functions~\cite{exponentdef,maxarea1,exponent}.
    
    \section{Joule-Thomson expansion}\label{sec4}
    In 1852, J. P. Joule and W. Thomson conducted a porous plug experiment to study the internal energy of gases, leading to the discovery of the Joule-Thomson effect. In their  experiment, the gas undergoes an adiabatic throttling process when it flows from high pressure to low pressure, resulting in a change in temperature. This process is isenthalpic:
    \begin{equation}\label{eq4_1}
    	\mathrm{d}H=0.
    \end{equation}
    The Joule-Thomson coefficient is defined as
    \begin{equation}\label{eq4_2}
    	\mu=\left(\frac{\partial T}{\partial P}\right)_{H}.
    \end{equation}
    The regions where $\mu>0$ and $\mu<0$ are the cooling region and the heating region respectively. The point where $\mu=0$ is referred to as the inversion point.
     
    The Joule-Thomson expansion for AdS black holes in the extended phase space has been widely studied~\cite{JTRN,JTKerr,JTBI,JTBD1,JTBD2,JThigh1,JThigh2,JThigh3}. Since it is suggested that the black hole's mass $M$ is equivalent to its enthalpy~\cite{MPfirstlaw}, the Joule-Thomson expansion for AdS black holes corresponds to a constant mass process $\mathrm{d}M=0$.
    
\begin{figure}[htbp]
    	\centering
    	\includegraphics[width=0.6\textwidth]{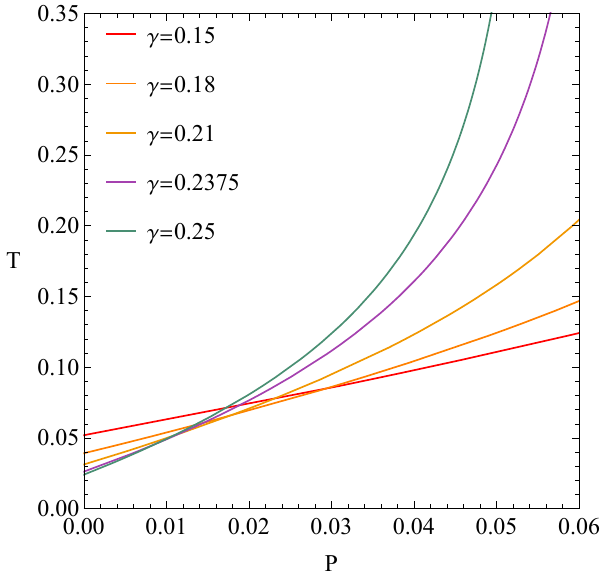}
    	\caption{The inversion curves for different values of the Barbero-Immirzi parameter $\gamma$. We set $\gamma=0.15, 0.18, 0.21, 0.2375, 0.25$ in five curves respectively.}\label{inversioncurve}
    \end{figure}

    Eq.~\ref{eq3_a1}, Eq.~\ref{eq3_a2} and the negative branch of Eq.~\ref{eq2_4} give
    \begin{equation}\label{eq4_3}
    	P\left(\alpha,M,r_{+}\right)=\frac{-3\alpha M^{2}+6Mr_{+}^{3}-3r_{+}^{4}}{8\pi r_{+}^{6}},
    \end{equation} 
    \begin{equation}\label{eq4_4}
    	T\left(\alpha,M,r_{+}\right)=\frac{-3\alpha M^{2}+3Mr_{+}^{3}-r_{+}^{4}}{2\pi r_{+}^{5}},
    \end{equation}
     with the minimum radius of event horizon for a fixed mass $M$:
    \begin{equation}\label{eq4_5}
    	r_{+}^{\mathrm{min}}=\sqrt[3]{\alpha M}.
    \end{equation}
    The Joule-Thomson coefficient is
    \begin{equation}\label{eq4_6}
    	\mu=\left(\frac{\partial T}{\partial P}\right)_{M}=\left( \frac{\partial T/\partial r_{+}}{\partial P/\partial r_{+}}\right)_{M}=\frac{30\alpha M^{2}r_{+}-12Mr_{+}^{4}+2r_{+}^{5}}{9\alpha M^{2}-9Mr_{+}^{3}+3r_{+}^{4}}.
    \end{equation}
 	By setting $\mu$ to zero, one could obtain the inversion curves, which is presented in Fig.~\ref{inversioncurve}.
     
    For an inversion curve, it divides the $\left(P,T\right)$ phase space into two zones: the lower right side of the curve is the heating region, and the upper left side represents the cooling region. We will examine this in the constant mass curves later.
     
\begin{figure}[htbp]
    	\centering
    	\includegraphics[width=0.8\textwidth]{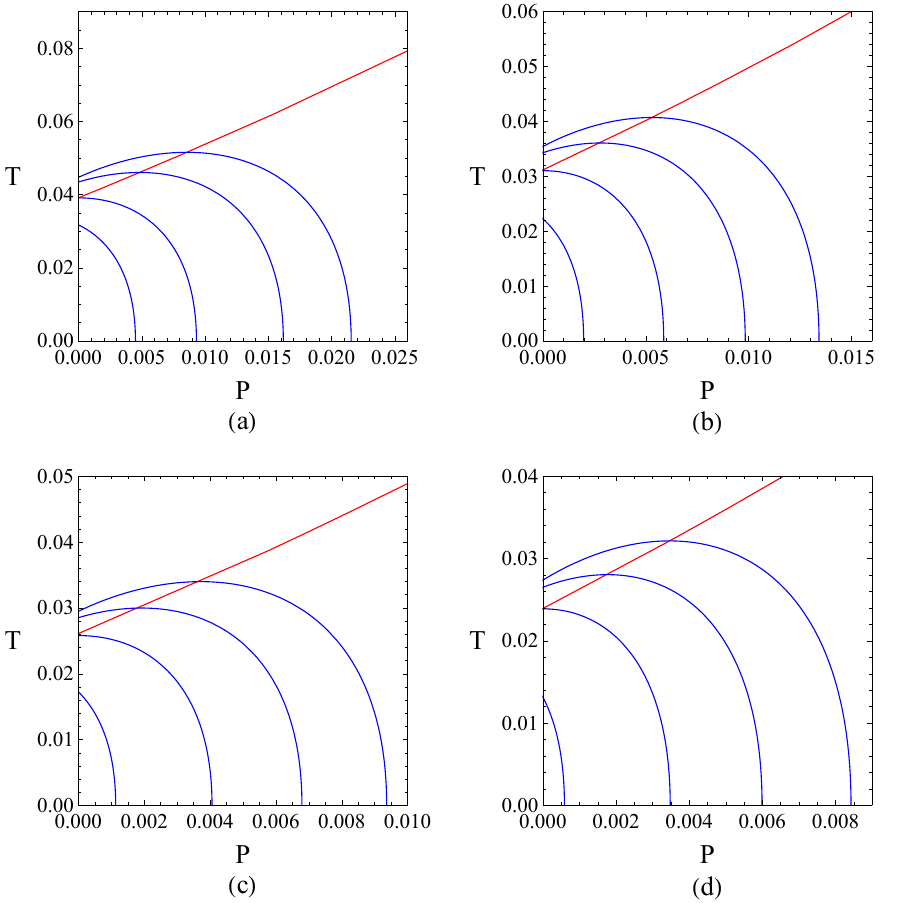}
    	\caption{The inversion curves and the constant mass curves for AdS-Schwarzschild-like black hole in LQG for different $\gamma$. The red lines represent the inversion curves and the blue lines are the isenthalpic lines. We set (a) $\gamma=0.18$, $M=0.57, 0.594, 0.63, 0.66$; (b) $\gamma=0.21$, $M=0.71, 0.75, 0.79, 0.83$; (c) $\gamma=0.2375$, $M=0.85, 0.90, 0.95, 1.0$; (d) $\gamma=0.25$, $M=0.91, 0.97, 1.03, 1.09$.}\label{constantM}
    \end{figure}

    For a fixed $M$, Equation $\mu=0$ has a root if and only if
    \begin{equation}\label{eq4_7}
    	M\geq M_{i}^{\mathrm{min}}=\frac{49\sqrt{6\alpha}}{144}.
    \end{equation}
    $M_{i}^{\mathrm{min}}$ is called the minimum inversion mass. Any black holes with a mass smaller than this value will not exhibit an inversion point. From the inversion curves, the inversion temperature will achieve its minimum when the inversion pressure $P_{i}=0$, leading to the expression
    \begin{equation}\label{eq4_8}
    	T_{i}^{\mathrm{min}}=\frac{\sqrt{6}}{28\pi\sqrt{\alpha}}.
    \end{equation}
    The ratio between the minimum inversion temperature and the critical temperature is
    \begin{equation}\label{eq4_9}
    	\frac{T_{i}^{\mathrm{min}}}{T_{c}}=\frac{5\sqrt{30}}{42},
    \end{equation}
    which also is a constant that is independent of $\gamma$. The inversion curves and the isenthalpic process for different parameters are shown in Fig.~\ref{constantM}.
    
    In figure, the mass $M$ of the constant mass curves increases from the inside to the outside. For each subfigure, the mass $M$ for the second isenthalpic curve is chosen as the minimum inversion mass. It is clear that the constant mass curve with the critical mass just reaches its inversion point at $P_{i}=0$. And any black holes with $M<M_{i}^{\mathrm{min}}$ are always in the heating process.
    
    \section{Conclusion and outlook}\label{sec5}
    We obtain the Schwarzschild-AdS space-time in LQG correction under the assumption that the cosmological constant is decoupled in LQG, and we studied its thermodynamics, including the modified first law, the equation of state and criticality. We drew $P-v$ graph and found that due to the LQG correction, the Schwarzschild-AdS black hole exhibits a critical point, and a critical ratio $7/18$, which slightly deviates from that of the Van der Waals system. The deviation of the critical ratio from $3/8$ is not uncommon in black hole thermodynamics, potentially introducing a new insight into the thermodynamics of AdS black holes and even into the studies of real gas system. Future astronomical observations or thermodynamic studies of real gases may reveal evidence supporting these findings. Moreover, it is concluded that if the energy-momentum tensor is related to the black hole's mass, the conventional first law does not hold. The corrected first law will violate the conservation of the Gibbs free energy during the phase transition, potentially indicating the occurrence of the zeroth-order phase transition. Furthermore, the Joule-Thomson expansion is also discussed in detail. We derived the inversion curves and displayed the isenthalpic process in $\left(P,T\right)$ graph, showing that the inversion curve will divide $\left(P,T\right)$ phase space into two regions: the heating region and the cooling region. Additionally, there is a minimum inversion mass, below which any black hole will always remain in the heating process. Since $\gamma=0.2375$ has been widely used~\cite{LQGmetric,LQGimage,LQGshadow,gammavalue3}, here we provide a brief list of important thermodynamic values in Tab.~\ref{tab} for reference. It is our hope that this research could provide a reference for the studies on AdS-black holes in LQG.
    \begin{table}[htbp]
        	\centering
       		\caption{List of important values in thermodynamics. All results are retained with six significant digits and $\gamma=0.2375$. Critical ratio and $T_{i}^{\mathrm{min}}/T_{c}$ are independent of $\gamma$.}
        	\includegraphics[width=0.8\textwidth]{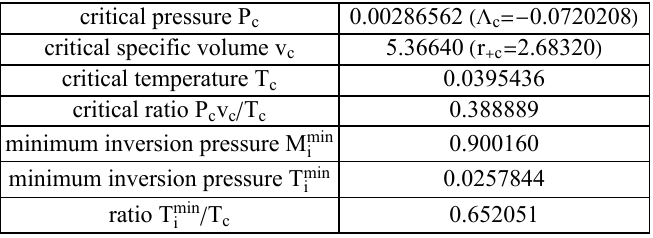}\label{tab}
    \end{table}
        
    It is necessary to claim that in this article we did not consider the coupling between LQG and cosmological constant. We believe that taking coupling effect into account will be an interesting subject. On the other hand, due to the conventional first law holds no longer, differential forms $T\mathrm{d}S+V\mathrm{d}P$ and $-S\mathrm{d}T+V\mathrm{d}P$ are not exact forms. Perhaps there will be a better way to define black hole's enthalpy and Gibbs free energy. Moreover, thermodynamics of rotating black hole is also studied in~\cite{Dolan,rotate1,rotate2,rotate3,rotate4}. So it will bring us more interesting phenomena if we consider the rotation of black hole in LQG.

	\section*{Conflicts of interest}
    The authors declare that there are no conflicts of interest regarding the publication of this paper.
    
    \section*{Acknowledgments}
    We want to thank the National Natural Science Foundation of China (Grant No.11571342) for supporting us on this work.
	\appendix
	\section{Modified first law}\label{app1}
	\setcounter{equation}{0}
	\renewcommand{\theequation}{A.\arabic{equation}}
	We denote the metric function of the black hole as $f\left(M,r\right)$, where $M$ is black hole's mass, and the radius of horizon $r_{+}$ ensures
	\begin{equation}\label{aeq1}
		f\left(M,r_{+}\right)=0.
	\end{equation}
	Taking the derivative of $r_{+}$ for both sides, one have
	\begin{equation}\label{aeq2}
		\frac{\partial f}{\partial M}\frac{\partial M}{\partial r_{+}}+\frac{\partial f}{\partial r_{+}}=0.
	\end{equation}
	The temperature $T_{k}$ defined by black hole's surface gravity is
	\begin{equation}\label{aeq3}
		T_{k}=\frac{1}{4\pi}\frac{\partial f}{\partial r_{+}}.
	\end{equation}
	The temperature $T_{H}$ derived from the conventional first law $\mathrm{d}M=T_{H} \mathrm{d}S+V \mathrm{d}P$ is
	\begin{equation}\label{aeq4}
		T_{H}=\frac{\partial M}{\partial S}=\frac{1}{2\pi r_{+}}\frac{\partial M}{\partial r_{+}}.
	\end{equation}
	By using the modified first law $W\left(r_{+},\Lambda\right)\mathrm{d}M=T_{k} \mathrm{d}S+V\mathrm{d}P$, it is clear that
	\begin{equation}\label{aeq5}
		W=\frac{T_{k}}{T_{H}}.
	\end{equation}
	Using Eq.~\ref{aeq2}, Eq.~\ref{aeq3} and Eq.~\ref{aeq4}, one could derive
	\begin{equation}\label{aeq6}
		W=-\frac{r_{+}}{2}\frac{\partial f}{\partial M}\Big|_{r_{+}}.
	\end{equation}
	This formula lacks physical significance. We need to rewrite it in another form. By using the Einstein equation Eq.~\ref{eq3_a7}, one get
	\begin{equation}\label{aeq7}
		\frac{\partial \left(r f\right)}{\partial r}=8\pi r^{2} T_{0}^{0}-r^{2}\Lambda +1.
	\end{equation}
	Taking the derivative of $M$ on both sides, one could get
	\begin{equation}\label{aeq8}
		\frac{\partial}{\partial r}\left(r \frac{\partial f}{\partial M}\right)=8\pi r^{2} \frac{\partial T_{0}^{0}}{\partial M}.
	\end{equation}
	Integrating both sides with respect to $r$ from $r_{+}$ to $+\infty$, one could obtain
	\begin{equation}\label{aeq9}
		\int_{r_{+}}^{+\infty}\frac{\partial}{\partial r}\left(r \frac{\partial f}{\partial M}\right)\mathrm{d}r=	\int_{r_{+}}^{+\infty}8\pi r^{2} \frac{\partial T_{0}^{0}}{\partial M}\mathrm{d}r.
	\end{equation}
	One could derive
	\begin{equation}\label{aeq10}
		W=-\frac{r_{+}}{2}\frac{\partial f}{\partial M}=\int_{r_{+}}^{+\infty}4\pi r^{2} \frac{\partial T_{0}^{0}}{\partial M}\mathrm{d}r+\varphi,
	\end{equation}
	where 
	\begin{equation}\label{aeq11}
		\varphi=\lim_{r\to \infty}\left(-\frac{r}{2}\frac{\partial f}{\partial M}\right).
	\end{equation}
	Here we assume that the metric of black hole asymptotically approaches the Schwarzschild-AdS spacetime at infinity. This means
	\begin{equation}\label{aeq12}
		f=1-\frac{2M}{r}-\frac{\Lambda r^{2}}{3}+\mathcal{O}\left(r^{-2}\right),\qquad r\to+\infty.
	\end{equation}
	Using this condition, it is easy to verify
	\begin{equation}\label{aeq13}
		\varphi=1.
	\end{equation}
	Eq.~\ref{eq3_a6} has been proven. This conclusion reveals that if the energy-momentum tensor of spacetime contains the black hole's mass, the conventional first law will be broken. 
	
	Please notice that the assumption Eq.~\ref{aeq12} is necessary, as it could be proven that even a slight deformation of the Schwarzschild-AdS spacetime would invalidate this conclusion. Considering this metric
	\begin{equation}\label{aeq14}
		f=1-\frac{\left(2+\epsilon\right)M}{r}-\frac{\Lambda r^{2}}{3},\qquad \epsilon\ll 1,
	\end{equation}
	which results in a correction function $W=1+\epsilon/2$. However, in this case, the energy-momentum tensor is still zero.

   	\bibliographystyle{unsrt}
	\bibliography{paper}
\end{document}